# SBML2Julia: interfacing SBML with efficient nonlinear Julia modelling and solution tools for parameter optimization

Paul F. Lang[1,*], Sungho Shin[2], Victor M. Zavala[2]

[1]Department of Biochemistry, University of Oxford, South Parks Road, Oxford OX1 3QU, UK.

[2]Department of Chemical and Biological Engineering, University of Wisconsin-Madison, Madison, WI 53706, USA

[*]To whom correspondence should be addressed.

**Abstract**

**Motivation:** Estimating model parameters from experimental observations is one of the key challenges in systems biology and can be computationally very expensive. While the Julia programming language was recently developed as a high-level and high-performance language for scientific computing, systems biologists have only started to realise its potential. For instance, we have recently used Julia to cut down the optimization time of a microbial community model by a factor of 140. To facilitate access of the systems biology community to the efficient nonlinear solvers used for this optimisation, we developed SBML2Julia. SBML2Julia translates optimisation problems specified in SBML and TSV files (PEtab format) into Julia for Mathematical Programming (JuMP), executes the optimization and returns the results in tabular format.

**Availability and implementation:** SBML2Julia is freely available under the MIT license. It comes with a command line interface and Python API. Internally, SBML2Julia calls the Julia LTS release v1.0.5 for optimisation. All necessary dependencies can be pulled from Docker Hub (https://hub.docker.com/repository/docker/paulflang/sbml2julia). Source code and documentation are available at https://github.com/paulflang/SBML2Julia.

**Contact:** paul.lang@wolfson.ox.ac.uk

## 1 Background

Biological systems frequently involve highly dynamic interactions between their constituent components. These interactions are often described in ordinary differential equation (ODE)-based models. Such models comprise constant parameters whose value is typically determined by fitting the model to experimentally observed data. This optimisation task is complicated by the high dimensionality of the optimisation problem, the existence of several local optima, flat areas in the objective function and structural and practical parameter unidentifiability (Hass *et al.*, 2019; Raue *et al.*, 2009; Villaverde *et al.*, 2019; Gábor and Banga, 2015). Most optimization algorithms for biological problems utilize strategies that iterate through multiple steps of model simulation and parameter adjustment (Raue *et al.*, 2015; Balsa-Canto *et al.*, 2016; Schälte *et al.*, 2020). Alternatively, such optimisation task can be treated as a dynamic optimisation problem, which has been extensively studied in engineering disciplines (Biegler, 2010). Dynamic optimisation is a class of optimisation problems with embedded dynamical systems. Such problems can be transformed into finite-dimensional nonlinear optimisation problems by discretising the ODEs and directly embedding them as equality constraints for the optimisation. This procedure is called direct transcription method and yields a large-scale optimisation problem. However, by exploiting sparsity the problem can be solved efficiently. The implementation of such procedure is facilitated by the algebraic modelling tool JuMP (Dunning *et al.*, 2017) and generic nonlinear optimisation solver Ipopt (Wächter and Biegler, 2006), both of which are available in the general-purpose programming language Julia (Bezanson *et al.*, 2017). Shin et al. have recently demonstrated the power of this approach by cutting down optimisation time of a microbial community model by a factor of 140 (Shin *et al.*, 2019). However, their code was custom made for the specific microbial community model.

Systems biology models are typically stored and exchanged in the Systems Biology Markup Language (SBML) format, which has a very broad range of software support. Schmiester et al. have recently developed the PEtab format which extends upon SBML model specification, by providing detailed information for parameter optimisation such as experimental conditions, measurement values and observable quantities and parameter properties in a structured form as TSV files (Schmiester *et al.*, 2020). As of today, PEtab is partially supported by 8 biological optimisation tools.

We developed SBML2Julia to provide the systems biology community with an easy to use interface to the highly efficient optimization tools available in Julia/JuMP (Fig. 1). SBML2Julia translates a PEtab specified optimisation problem into a JuMP model written in Julia language, performs the optimisation using the nonlinear optimization solver Ipopt, and returns the results in tabular format. Moreover, the direct-transcription method based optimization enables (a) optimisation without providing initial conditions as extra parameters or results of costly pre-equilibration, (b) simulation-free (constraint-based) pre-equilibration and (c) exploiting highly efficient and robust automatic differentiation tools, filter line-search interior-point solvers, and sparse linear algebra routines.

**Python interface:**
```
>>> import SBML2Julia

>>> problem = SBML2Julia.SBML2JuliaProblem('petab_problem.yaml')
>>> problem.optimize()
>>> res = problem.results
```
**Command line interface:**
```
user@bash:/$ SBML2Julia optimize 'petab_problem.yaml' -d './results/'
```

**Figure 1: SBML2Julia user interfaces.** SBML2Julia provides easy to use Python and command line interfaces to efficient nonlinear optimization tools in Julia/JuMP. Users only need to specify the path to the PEtab problem specification YAML file (part of the PEtab standard, contains paths to the SBML model and TSV condition, observable, parameter, and measurement files). The `-d` flag in the command line interface allows specification of the output/results directory.

## 2 Features

SBML2Julia offers users the following features:

- Complete PEtab support: basic time-course calculation, multiple experimental conditions, numeric observable parameter overrides in measurement table, parametric observable parameter overrides in measurement table, parametric overrides in condition table, time-point specific overrides in the measurement table, observable transformations to log10 scale, replicate measurements, pre-equilibration, partial pre-equilibration, numeric initial concentration in condition table, numeric initial compartment sizes in condition table, parametric initial concentrations in condition table, numeric noise parameter overrides in measurement table, parametric noise parameter overrides in measurement table, observable transformations to log scale, normal and Laplacian distributed noise type, optimisation of noise parameters, regularization with (log-)normal and (log-)Laplacian parameter priors.
- TSV and Excel export containing a workbook with the best parameter set, time-courses of all species, time-courses of all observables, the objective function value as negative logarithm of the posterior probability of the parameters given the data and the Chi2 value of the residuals.

- TSV export of the optimal solution in PEtab format for interoperability with other PEtab supporting tools.
- Plots of calculated ("simulated") and measured observables.
- Designed for, but not limited to biological optimisation problems.
- Multi-start optimisation: theoretically the interior point algorithm underlying SBML2Julia can get stuck in local minima (Shin *et al.*, 2019). Using multi-start optimisation, we have empirically found that our problems frequently converge to the same solution regardless of the starting point.
- Customizable number of time-discretisation steps to check for potential discretisation errors.
- Output of internally translated Julia/JuMP code.
- Optional and flexible modification of internally translated Julia/JuMP code: gives users maximal flexibility to modify the optimization problem.
- Initial condition-free optimization: fitting ODE models requires adding initial conditions as extra parameters subject to optimization or pre-equilibration. Due to the time-discretization in SBML2Julia providing initial conditions is optional (constitutes a further optimization constraint).
- Simulation-free pre-equilibration: rather than performing time consuming pre-equilibration simulations, SBML2Julia implements pre-equilibration based on the constraint that $d\mathbf{x}/dt(t=0, \mathbf{c_p}) := 0$, where $\mathbf{x}$ are the species and $\mathbf{c_p}$ is the pre-equilibration condition.
- A Docker image for simple installation of dependencies.
- Easy installation from GitHub or PyPI.
- Open source with a permissive MIT license.
- Support for the third party HSL library (HSL developers.) of linear solvers.

## 3 Examples and Documentation

SBML2Julia comes with extensive documentation and tutorials demonstrating its functions. We further provide two example optimization problems, one with synthetic data, one with real experimental measurements. The synthetic problem is based on a G2/M cell cycle transition model (Vinod and Novak, 2015). It contains 13 species, 13 observables, 24 parameters and 4 "experimental" conditions. The real data problem is a gut microbiome community model (Shin *et al.*, 2019). In the implemented PEtab formulation it contains 2 species, 2 observables, 156 parameters and 210 experimental conditions.

## 4 Conclusion

SBML2Julia provides the systems biology community with an easy to use and flexible command line and Python interface between SBML models and the highly efficient non-linear optimisation solvers for parameter estimation. The a priori time-discretization adds a completely new optimisation strategy to the current PEtab ecosystem. We hope this will help researchers to find the ideal method for solving their optimisation problems without going through the painstaking process of constantly adapting the problem specification to the method of interest.

## Funding

Paul F Lang acknowledges funding from the University of Oxford and the EPSRC & BBSRC Centre for Doctoral Training in Synthetic Biology (grant EP/L016494/1).